\documentclass[aps,prl,reprint,amsmath,amssymb, showkeys, superscriptaddress]{revtex4-1}
\usepackage[T1]{fontenc}
\usepackage{graphicx}
\usepackage{dcolumn}
\usepackage{bm}
\usepackage{breqn}
\usepackage{color}
\makeatletter
\let\cat@comma@active\@empty
\makeatother
\usepackage{setspace}
\usepackage[footnotesize]{caption}
\begin{document}

\title{Ring statistics in  2D-silica: effective temperatures in equilibrium}

\author{Projesh Kumar Roy}
\affiliation{NRW Graduate School of Chemistry, Wilhelm-Klemm-Stra{\ss}e 10, 48149 M\"unster, Germany}
\affiliation{Institute f\"ur Physikalische Chemie, Westf\"alische-Wilhelms-Universit\"at M\"unster, Corrensstra{\ss}e 28/30, 48149 M\"unster, Germany}

\author{Andreas Heuer}
\email{andheuer@uni-muenster.de}
\homepage{https://www.uni-muenster.de/Chemie.pc/heuer/}
\affiliation{Institute f\"ur Physikalische Chemie, Westf\"alische-Wilhelms-Universit\"at M\"unster, Corrensstra{\ss}e 28/30, 48149 M\"unster, Germany}

\date{\today}



\begin{abstract}

The thermodynamic properties of subsystems in strong interaction with the neighborhood can largely differ from the standard behavior. Here we study the thermodynamic properties of rings and triplets in equilibrated disordered 2D-silica. Their statistics follows a Boltzmann behavior, albeit with a strongly reduced temperature. This effective temperature strongly depends on the length scale of the chosen subsystem.  From a systematic analysis of the 1D Ising model and an analytically solvable model we suggest that these observations reflect the presence of strong local positive energy correlations.  

\end{abstract}

\maketitle

\begin{spacing}{1.1}

\textbf{Introduction}--- The thermodynamic properties of large systems in a heat bath are very well understood in the framework of the canonical ensemble.   However, the situation becomes more complex when small systems such as molecules in solution are analyzed. One expects strong coupling effects, which may give rise to new phenomena \cite{Hill2002}.  As a very clear-cut example Dixit has studied the properties of a dumbbell in a solvent with respect to the distance distribution along the internal harmonic coordinate \cite{DixitJChemPhys2013, DixitPCCP2015}. Interestingly, this distribution showed significant deviations from the standard Boltzmann behavior, expected for thermal equilibrium. However, the average energy still allows one to read of the temperature so that a dumbbell can still be regarded as a nanothermometer \cite{VulpianiAmJPhys2011}. The numerical data could be explained in the framework of a hyperensemble  \cite{BeckPhysicaA2003, CrooksPhysRevE2007} which can be expressed as a distribution of temperatures and can be directly related to non-extensive statistics \cite{TsallisJStatPhys1988}.  Other researchers have used the concept of a temperature distribution in order to rationalize the emergence of dynamic heterogeneities in glass-forming systems \cite{ChamberlinEntropy2015}. In any event, it is not obvious whether the distribution of temperatures is a useful concept at all \cite{KittelPhysToday1988, VulpianiAmJPhys2011}. Also fluctuations of other intensive thermodynamic properties such as the chemical potential have been suggested to describe local structural phenomena \cite{BansalJChemPhys2016, BansalJChemPhys2017, DixitJChemPhys2017}.

Here we discuss a realistic system of an atomic disordered network. The recent discovery of the two-dimensional silica networks (2D-silica) with STM \cite{HeydeAngew2012, HeydeJPhysChemC2012} /STEM \cite{HuangNanoLett2012} opened up the possibility to compare the simulated and the experimentally analyzed network on the level of individual atoms \cite{RoyPCCP2018}. Recently, we have developed a two-dimensional Yukawa type force-field \cite{AlejandreJChemPhys2005, AlejandreCondMattPhys2012, AlejandreJChemPhys2012} to quantitatively describe the structure formation of 2D-silica (for details, see Ref. \citenum{RoyPCCP2018}). With this model it was possible to reproduce many experimental observations, related to the structural properties, as well as the distributions of rings and `triplet of rings'. In general, such systems can be discussed in the framework of random networks which was mainly developed from the study of various macroscopic objects present in nature, e.g.,  soap bubbles or plant cells \cite{RivierPhilMagB1985}. Using Jaynes's interpretation of entropy in information theory \cite{ShannonBellSysTechJ1948, JaynesPhysRev1957}, Rivier et al. \cite{RivierJPhysA1982, RivierPhilMagB1985} postulated a `maximum entropy formulation' of random networks which can predict the occurrence probabilities  of different ring sizes based on geometric parameters.

A natural elementary unit of a random network is a ring. One may check the correlation between their energies and their observation probabilities in equilibrium. However, as compared to the properties of small molecular units in solution, there is one conceptual difference: every atom belongs to more than one ring. Thus, the interesting question emerges whether the resulting interaction between a ring and its surrounding gives rise to small-system effects, like the  non-Boltzmann type distribution observed for the dumbbell in solution.

Our key goal of this work is to show numerically that the probability of a ring is governed by its energy via an approximate Boltzmann relation, albeit with a largely different temperature. Similar observations are made when applying a reweighting procedure or studying substructures of the 1D Ising model.  From a model analysis we relate this observation to the presence of positive energetic correlations of adjacent regions of the system. Simulations and model analysis show consistently that the standard thermodynamic behavior is approached upon increasing the size of the analysed substructures.

\textbf{Energetic aspects}--- The Yukawa type force-field to simulate the network formation of 2D-silica in two-dimensions reads

\begin{equation}
  V_{ij}(r_{ij}) = [(\sigma_{ij}/r_{ij})^{12} + (q_{ij}/r_{ij}) \exp(-\kappa r_{ij})]
  \label{eqn:yukawa}
\end{equation}

where the force-field parameters ($\sigma,q,\kappa$) are given in Ref. \cite{RoyPCCP2018}. We have performed long NVT simulations with the Nose-Hoover thermostat \cite{NoseJChemPhys1984} with a time-step 0.01 for $T \leq 0.016$ and 0.005 for higher temperatures. We restricted our analysis to long times where equilibrium was reached. After the simulation was complete, we locally minimized the evenly stored configurations of the system with respect to its energy to sample the underlying inherent \textcolor{red}{structures}, i.e. removing the vibrational fluctuations \cite{StillingerPhysRevA1982,HeuerJPhysCondensMatt2008}. We filtered out all `defect' configurations where the coordination criteria for silicon and oxygen in two-dimensions are not fulfilled for each particle \cite{RoyPCCP2018}. For the `defect-free' configurations each structure is composed of 16 rings, covering the entire area without any gap in connectivity. It is essential to use small systems because otherwise one hardly finds defect-free configurations. It was checked that the ring statistics does not display any relevant finite size effects \cite{RoyPCCP2018}.

\begin{center}
\includegraphics[height=4cm,width=8cm]{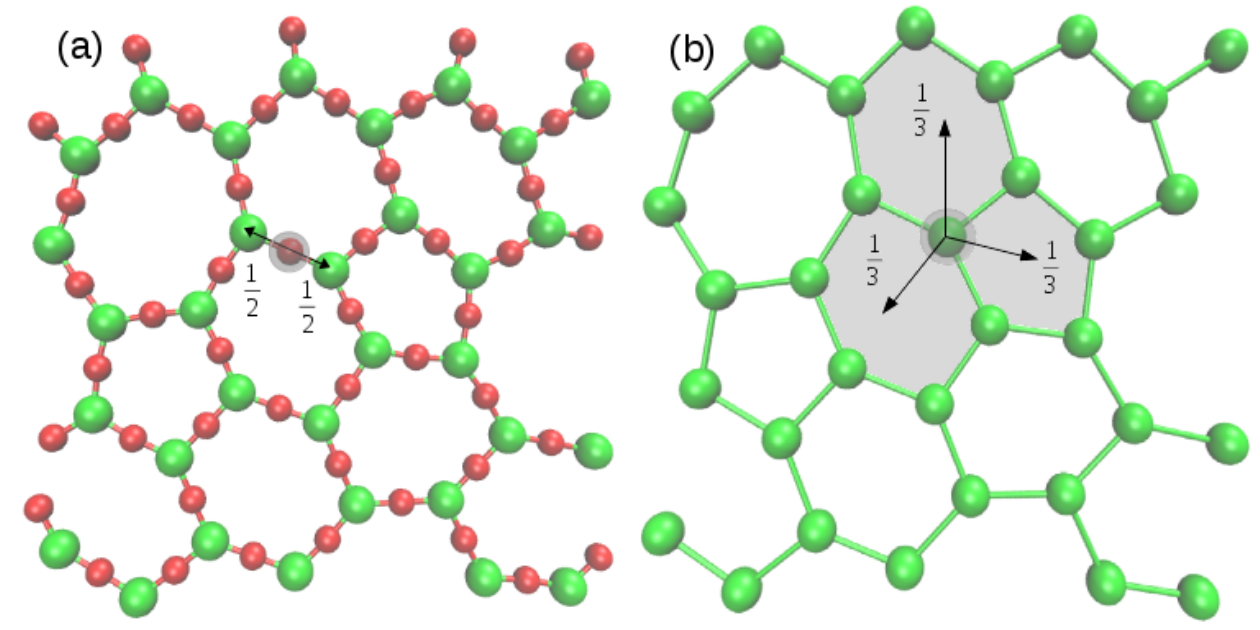}
\captionof{figure}{Sketch of the definition of the ring energies for a defect-free configuration, using periodic boundary conditions. (Red = O, Green = Si) . (a): All O-particle energies are equally added to the connected Si-particles. (b): All redefined Si-particle energies are equally redistributed to the connected rings. The combination of three rings in (b), marked with gray shade, is an example of a `triplet of rings' as they share a common Si particle at the center. The image is partially reproduced from Ref. \citenum{RoyPCCP2018} by permission of the PCCP Owner Societies.}
\label{fig:onethirdrule}
\end{center}

A natural definition of ring energies, based on the energies of the individual atoms, is sketched in Fig.\ref{fig:onethirdrule}. In a first step the potential energy of every O particle is equally partitioned to the two adjacent Si particles; see Fig. \ref{fig:onethirdrule}. As a result, the Si particles can be regarded as `effective' Si particles since they also contain information about the neighboring O particles. Formally this can be written as

$$\epsilon_{Si,eff} = \frac{1}{2}[ \epsilon_{Si} + \frac{1}{2}\sum_{i = 1}^3 \epsilon_{O}^i ]$$

where the sum is over all three oxygen neighbors of a silicon particle. The pre-factor $\frac{1}{2}$ takes into account the fact that for the employed two-body force fields each energetic contribution appears twice. In the final step the energies of all effective Si particles are equally distributed to the three rings, they belong to. In what follows we denote the effective Si particles as `particles'. The individual ring energy is denoted $\epsilon_r$. Furthermore, we define $E_r$ as the average ring energy of rings of size $r$. With this construction the sum of the ring energies is identical to the total energy of the system.

Further, we analyze triplets of rings \cite{HeydeJnonCrysSolid2016}, formed by the three rings with a common central-corner. Their energies or sizes are just the sum of the energies or sizes of the contributing rings. The average energy of a triplet, $E_t$, is defined in analogy to $E_r$. For the detailed data, see SI.I, SI.II.

\textbf{Boltzmann analysis}---  A defect-free 2D-silica network with periodic boundary conditions strictly conserves the average ring size of 6 \cite{RoyPCCP2018}. This additional constraint can be taken into account by an additional Lagrange parameter. However, in order to directly extract correlations between energies and probabilities, we use a simple trick. We define for a ring with size $r$ the `complementary' ring with size $r^* = 12 - r$ such that the average size of both rings is 6.  As a consequence, we can correlate the energy $E_r + E_r^*$ with the probability $P_r P_r^*$ without additional size constraints.

\begin{figure*}
 \includegraphics[width=18cm,height=6cm]{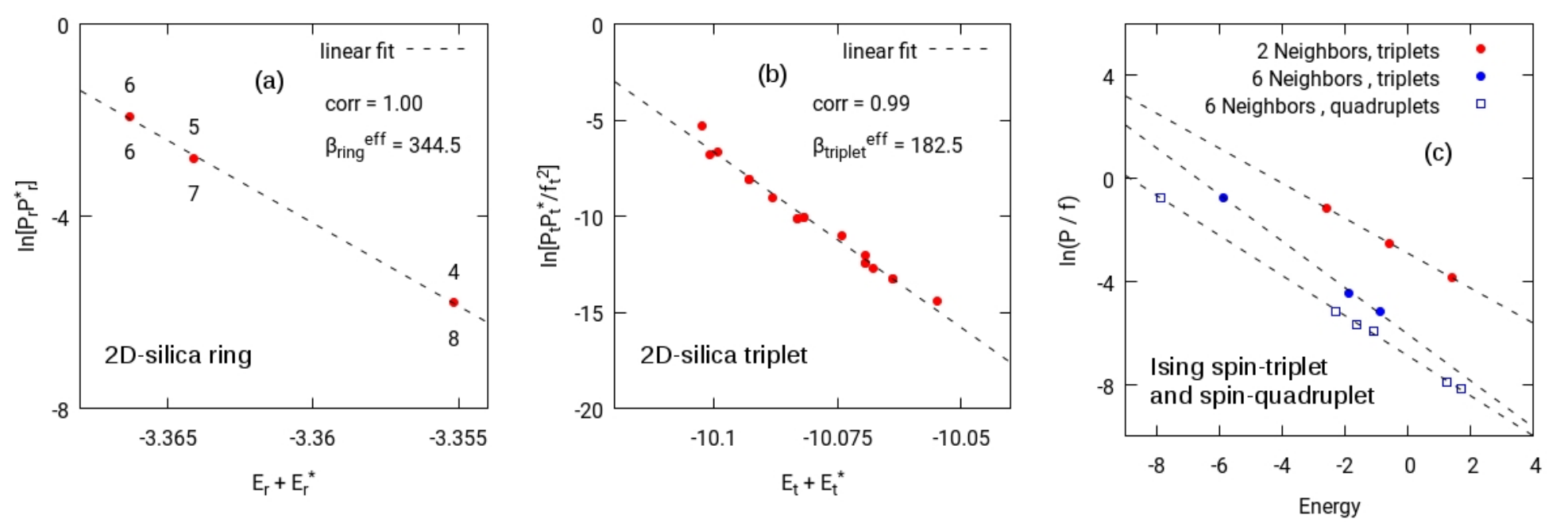}
 \caption{Comparison of the logarithm of (a) complementary ring probabilities and (b) complementary triplet probabilities at $T = 0.015$ ($\beta = 66.7$) of the 2D-Silica ring-network, with their respective average energies. The correlation coefficients are shown in the graph. For (a) and (b), the inverse effective temperatures are $\beta_{ring}^{eff} = 344.5$ and $\beta_{triplet}^{eff} = 182.5$, respectively. For (b), data is shown whenever $P_tP_t^* > 10^{-6}$.
 In (c) the probabilities of the different states of spin triplets and quadruplets for the 1D Ising model are compared with the respective average energies at $T = 1.5$ ($\beta = 0.67$).  The permutation factors for the substructures are represented by `f'.  The respective interaction range (2 or 6 neighbors) is listed in the figure.  The resulting inverse effective temperatures $\beta_{Ising}^{eff}$ from the Boltzmann fit are 0.67 ($\approx \beta$) (triplets, 2 neighbors),  0.90 (triplets, 6 neighbors) and 0.78  (quadruplets, 6 neighbors).}
 \label{fig:comp_energy}
\end{figure*}

We restrict this analysis to ring sizes between 4 and 8 since the other probabilities (in particular 3) are very small. For the temperature $T = 0.015$ one observes a nearly perfect Boltzmann behavior; see Fig.\ref{fig:comp_energy}(a). However, the required inverse temperature $\beta^{eff}_{ring}$ is much larger than $\beta$. We have repeated this analysis for different simulation temperatures, showing on average $\mu_{ring} \equiv \beta^{eff}_{ring}/\beta  \approx 5.1 $ (see SI.III). We checked for $T = 0.015$, using the temperature distribution as suggested by Dixit \cite{DixitJChemPhys2013, DixitPCCP2015}, that any residual temperature distribution is, if at all, very small. Interestingly, in the low-temperature limit the data start to deviate from a Boltzmann behavior and a temperature distribution might indeed become relevant (see SI.IV).

For triplets we proceed analogously. Again we avoid the use of additional Lagrange parameters to control the correct average size by introducing complementary triplets, e.g. 458 and 478. This guarantees that the average triplet size is 18, so that a direct correlation of $E_t + E_t^*$ with the probability $P_t P_t^*$ is possible.One has to use a degeneracy factor $f_t$ to take care of the different permutations \cite{RoyPCCP2018}.  As seen in Fig. \ref{fig:comp_energy}(b), we have an excellent Boltzmann relation with $\beta_{triplet}^{eff} > \beta$. Averaging  over the simulations at different temperatures, we obtain $ \mu_{triplet} \equiv \beta^{eff}_{triplet}/\beta  \approx 2.7$ (see SI.III).

How to interpret the values of $\mu_{ring}, \mu_{triplet}$? If the standard Boltzmann behavior would hold on the single particle level one would trivially obtain $\mu_{ring} = 3$, reflecting that in our definition of rings only one third of the particle energy is taken into account. Thus, to judge the amount of non-standard Boltzmann behavior one should take $\mu_{ring}^{eff} = 5.1/3 \approx 1.7$, corresponding to an increase of 70\% of the inverse bath temperature. Repeating the analysis for triplets, one has to take into account that on average a triplet contains 13 different particles. This results in $\mu_{triplet} = 13*3/18 \approx 2.2$ for the standard Boltzmann behavior. Then one obtains  $\mu_{triplet}^{eff} = 2.7/2.2\approx 1.2$ which is significantly smaller than $\mu_{ring}^{eff}$ and close to the value of unity for standard Boltzmann behavior. This agrees with the intuitive expectation that larger subsystems approach the standard thermodynamic behavior.

The same effect can be seen in an 1D Ising model with energy $E_{tot} = -\sum J_{ij}S_iS_j$ and $S_i \in {\pm 1}$. Here, we have compared the results for two different cases. In case A (standard Ising model), we choose $J_{ij}=1$ for two nearest neighbors and 0 otherwise. In case B, we choose the interaction strengths $J_{ij} = 1, 2/3, 1/3$ for both the first, second and third neighbor, respectively, and 0 otherwise. In analogy to the case of 2D-silica we first define the energy of a spin as half of the sum of all pair energies, related to that spin. Then, in analogy to the ring-triplets in 2D-silica, we define different subsystems via the orientation of three adjacent spins. Taking into account the symmetry-related cases, one ends up, e.g., with AAA, BAB and AAB configurations for the spin-triplets with degeneracy factors $f_t$ of 2, 2, and 4, respectively (see also SI.V). The analogous analysis can be also performed for spin quadruplets.

In a Boltzmann picture the probability to find a specific spin-triplet in equilibrium is given by $P_t/f_t \propto e^{-\beta E_t} $. Again, $E_t$ is the average energy of the respective triplet. In full analogy to 2D-silica we find a perfect Boltzmann behavior. Interestingly, for case B, one has to substitute $\beta$ by a larger inverse effective  temperature. This suggests that $\beta^{eff}/\beta$ is directly related to the interaction range and thus to the degree of spatially correlated behavior. For the spin-quadruplets, the inverse effective temperature is lower than for the spin-triplets, which again agrees with the observed size-dependence of $\mu$ for 2D-silica.

\begin{figure*}[!htb]
\centering
 \includegraphics[width=18cm,height=6cm]{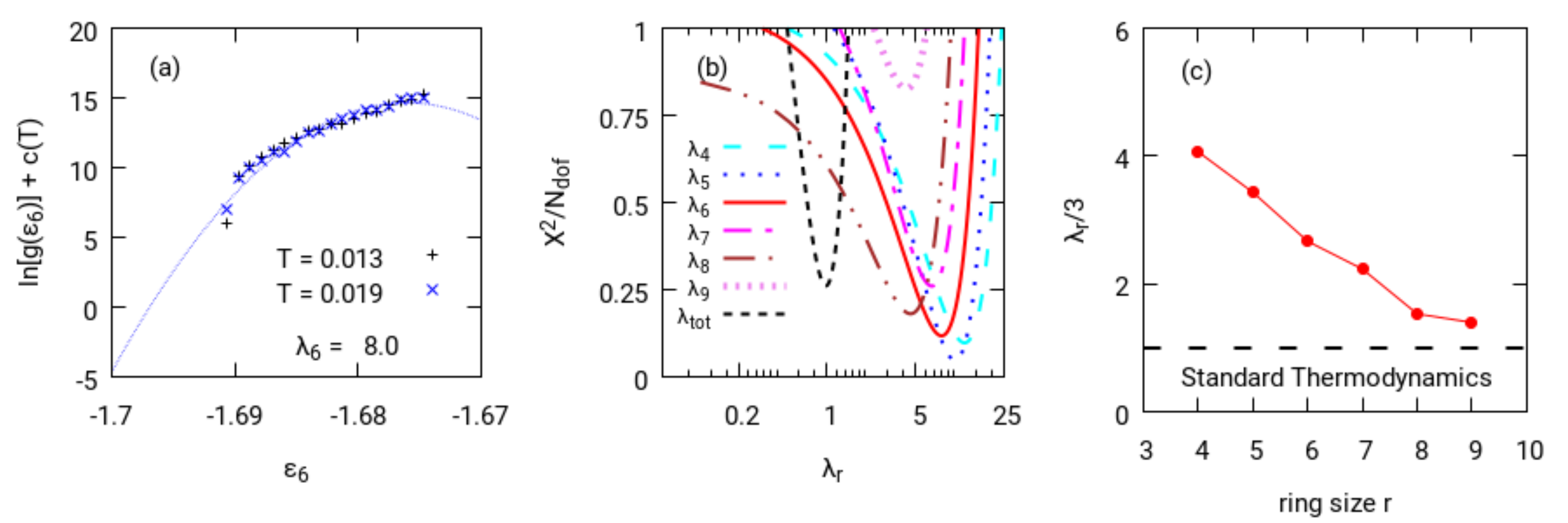}
 \caption{(a) Density of states for six member rings where the reweighting was performed at $\lambda_6 \beta$. The value of $\lambda_6=8.0$ is determined by minimizing the overlap error $\chi^2$ between the resulting density of states curves for $T = 0.013$ and $T = 0.019$. $c$ represents a simple shift parameter. We fit the density of states curves with a Gaussian function (see SI.VII for details about the Gaussian properties). All data are plotted where probabilities for both temperatures are larger than 0.002 (thus excluding the high-energy regime where anharmonic effects become relevant \cite{HeuerPhysRevE1999}). In (b), the dependence of $\chi^2/N_{dof}$ where the normalization factor denotes the number of data points minus the number of fitted parameters \cite{gnuplot}. The optimum $\lambda_{tot}$ value for the total defect-free system is very close to its theoretical value of 1.0. In (c), the size dependence of the optimum $\lambda_r$ values, scaled with the trivial factor of 3, are shown. For $r=9$ the noise is very high (see (b)) so that $\lambda_9$ should be taken with caution.}
 \label{fig:dos_6}
\end{figure*}

\textbf{Density of states analysis}--- So far we have analyzed the properties, e.g., of rings of a given size without resorting to the individual realization. In the next step, the configurations are studied individually. Previously, the distribution of minimized configurations (also denoted `inherent structures') of the total system has been analysed \cite{HeuerPhysRevLett2004, HeuerPhysRevE1999}. Their distribution results from the underlying  density of states $G(E_{tot})$, weighted by the Boltzmann factor. It is valid for sufficiently low temperatures and low energies such that the harmonic approximation holds \cite{HeuerPhysRevE1999,HeuerJPhysCondensMatt2000,HeuerJPhysCondensMatt2008}. Via Boltzmann reweighting, it is possible to extract the a priori unknown density of states $G(E_{tot})$, {\it independent} of the actually used temperature. Naturally, this procedure also works for the subensemble of defect-free states, sampled in equilibrium, as checked for the two extreme temperatures $T = 0.013$ and $T = 0.019$ (see Fig. \ref{fig:dos_6}(b) and SI.VI).

A similar procedure is now used for the distribution of rings of a given size $r$. The underlying distribution functions $g_r(\epsilon_r)$ are determined from Boltzmann reweighting of the equilibrium distributions of these two temperatures. We choose the inverse temperature factor as $\beta^{eff} = \lambda_r \beta$, and minimize the deviation between both reweighted curves, denoted $\chi^2(\lambda_r)$. As shown in Fig. \ref{fig:dos_6} for $r=6$ the optimum overlap is found for $\lambda_6 = 8.0$ which is much larger than the standard value of 3 as discussed above. Note that, except for $r=9$, the quality of the overlap, as derived from the minimum of $\chi^2(\lambda_r)$, is similar as for the reweighting of the total system. The optimum values of $\lambda_r$ decrease with increasing ring size (see Fig. \ref{fig:dos_6}(c) and SI.VII). Thus, in analogy to the above comparison of rings and triplets standard behavior is approached for larger units. We would like to remark that the value of $\mu_{ring} \approx 5$  is larger than typical values of $\lambda_r$ around 8 for the dominant ring sizes. It is satisfying that both approaches yield a significant increase of the inverse effective temperatures beyond the standard value of $3\beta$. A deeper comparison is difficult because in the first case one deals with the average energies of the different ring sizes whereas in the second case the individual energies of rings of a specific size are compared.

\textbf{Impact of additional correlations}--- If we identify the building blocks with the energies of the individual rings, one expects that adjacent rings are correlated in energy. A trivial reason is that the energy, related to individual atoms, are distributed among adjacent rings. Furthermore, it is known that, e.g., small rings like to be adjacent to large rings \cite{RoyPCCP2018}. One would also expect that the particle energies within a specific ring are correlated in order to fulfil the constraints to close that ring.

Here we show for a simple model system that correlations may give rise to the emergence of a size-dependent effective temperature in equilibrium situations.  We start by considering a system with two correlated identical subsystems $S_1$ and $S_2$ with energies $e_1$ and $e_2$. We assume that the density of states $g(e_i)$ of $S_i$ is Gaussian with variance $\sigma^2$. We define the correlation coefficient $\nu  = <e_1 e_2>/\sigma^2$. We require that the marginal densities of states are not modified by this correlation. It is easy to check that these conditions are fulled for the choice $e_2 = \nu e_1 + \epsilon_1 $ where $\epsilon_1 $ is a random number drawn from a Gaussian distribution with variance $\epsilon^2 = \sigma^2(1 - \nu^2)$. As a consequence the combined density of states $g(e_1, e_2)$ reads

\begin{equation}
g(e_1, e_2)  = \exp \left [-\frac{e_1^2}{2\sigma^2} \right ] \exp \left [-\frac{(e_2 - \nu e_1)^2}{2\epsilon^2} \right ]
\end{equation}

The probability of finding the energy doublet $(e_1, e_2)$ in a heat bath of inverse temperature $\beta$ can be written as $p(e_1, e_2) \propto g(e_1, e_2) \exp[-\beta(e_1 + e_2)]$. Integrating over $e_2$ one obtains the marginal distribution of $e_1$

\begin{equation}
p_1(e_1) \propto g(e_1) e^{-\beta[(1 + \nu)e_1]}
 \label{eqn:TSmodel_prob}
\end{equation}

Thus the resulting inverse effective temperature $\beta^{eff} = (1 + \nu) \beta$ depends on the degree of correlation. As shown in SI.VIII this analysis can be generalized to the case of $N$ subsystems where one analyses the energy distribution of units containing $M \le N$ subsystems. The above case corresponds to $N=2$ and $M=1$. One obtains $\beta^{eff}/\beta = 1 + (N-M)/(M - 1 + 1/\nu)$.Thus, upon increasing size M of the unit of interest standard behavior is approached. A possibly related size-dependence for fluctuations of the chemical potential is reported in \cite{DixitJChemPhys2017}.

\textbf{Discussion}--- In general one may expect deviations from standard thermodynamics if a subsystem is small and strongly interacting with the surrounding heat bath. For example, the dumbbell in a solvent displays a non-Boltzmann behavior where the average temperature, as deduced from the average harmonic energy, is identical to the solvent temperature. For the case of equilibrated 2D-silica we also see strong deviations from expectation. However, we still see an approximate Boltzmann behavior (with small deviations for the lowest temperatures), but with an effective temperature. Note that typically effective temperatures are used to describe non-equilibrium situations \cite{CarstenPRL2013, BarratJPhysCondMatt2003}.

For 2D-silica correlations between adjacent rings as well as within rings are expected. Thus, our general model, showing a connection of correlation and effective temperature, may indeed be relevant for the understanding of 2D-silica. In qualitative agreement with the model predictions we find that when either approaching the length scales of triplets or that of large rings nearly standard behavior is recovered. For the 1D Ising model we can recover these features, i.e. the dependence on the size of the unit of interest and the degree of correlations as tuned by the interaction range in that case. Thus, the specific deviations from standard thermodynamics, reported in this work, hold beyond 2D-silica. Note, however, that the model prediction $\beta^{eff}/\beta = const$ may be model specific (e.g., depending on the Gaussian properties) and is, e.g., violated for the 1D Ising model; see SI.III.

The key focus of the present work is the numerical identification of small-system thermodynamic effects for 2D-silica, which is an ideal model system to study small-scale thermodynamic properties due to the presence of well-defined local neighborhoods. Hopefully, it stimulates further analytical or numerical work to better understand, e.g., the emergence of apparent temperature distributions in the low-temperature limit, extending the non-standard Boltzmann behavior at higher temperatures, the dependence of the effective temperature on the ring size,  the quantitative relation to spatial ring correlations \cite{RoyPCCP2018, MahdiPhysRevE2016}, or the relation between $\mu$ and typical $\lambda_r$-values.  

\begin{acknowledgments}

We acknowledge support by the NRW Graduate School of Chemistry and the SFB 858  (DFG). We are grateful to R. V. Chamberlin and M. Heyde for important discussions.

\end{acknowledgments}

\end{spacing}

\bibliographystyle{apsrev4-1}

\providecommand{\noopsort}[1]{}\providecommand{\singleletter}[1]{#1}%

\end{document}